\pdfoutput=1

\documentclass[11pt]{article}

\usepackage{ACL2023}

\usepackage{times}
\usepackage{graphicx}
\usepackage{latexsym}

\usepackage[T1]{fontenc}

\usepackage[utf8]{inputenc}

\usepackage{microtype}

\usepackage{inconsolata}

\usepackage{graphicx}
\usepackage{amsmath} 

\newcommand{\tool}{{PatchRecall}~}
\newcommand{\toolnospace}{{PatchRecall}}
\title{\toolnospace: Patch-Driven Retrieval for Automated Program Repair}

\author{
Mahir Labib Dihan, Faria Binta Awal, Md. Ishrak Ahsan \\
Department of Computer Science and Engineering \\
Bangladesh University of Engineering and Technology (BUET) \\
\texttt{\{mahirlabibdihan, faria12mou, ishrak26\}@gmail.com}
}

\begin{document}
\maketitle
\begin{abstract}
Retrieving the correct set of files from a large codebase is a crucial step in Automated Program Repair (APR). High recall is necessary to ensure that the relevant files are included, but simply increasing the number of retrieved files introduces noise and degrades efficiency. To address this tradeoff, we propose \textbf{\toolnospace}, a hybrid retrieval approach that balances recall with conciseness. Our method combines two complementary strategies: (1) codebase retrieval, where the current issue description is matched against the codebase to surface potentially relevant files, and (2) history-based retrieval, where similar past issues are leveraged to identify edited files as candidate targets. Candidate files from both strategies are merged and reranked to produce the final retrieval set. Experiments on SWE-Bench demonstrate that \tool achieves higher recall without significantly increasing retrieved file count, enabling more effective APR.
\end{abstract}

\section{Introduction}
Modern software development relies heavily on collaborative platforms such as GitHub, where issues serve as central artifacts for tracking bugs, feature requests, and maintenance tasks. Large, popular repositories---for example, Django with more than 34K reported issues \cite{django30255} ---undergo rapid and extensive evolution. Addressing these issues often requires edits across multiple files and functions, making precise localization within a codebase both challenging and crucial \cite{DBLP:conf/issre/BissyandeLJRKT13}. 
Effective resolution of such issues depends on identifying the correct set of candidate files that need modification. 
However, navigating vast repositories exacerbates the tradeoff between recall and efficiency: retrieving too few files risks omitting relevant ones, while retrieving too many introduces noise and reduces efficiency.

Existing literature highlights several limitations in this space. 
Benchmarks such as HumanEval and MBPP \cite{austin2021program, chen2021evaluating} evaluate isolated and simplistic tasks that fail to capture the complexity of real-world repositories. 
Empirical studies demonstrate that Large Language Models (LLMs) struggle with repository-level tasks requiring class- and project-wide reasoning \cite{du2023classeval, liu2023lost}. 
To bridge this gap, SWE-bench \cite{jimenez2023swebench} was introduced, providing a benchmark derived directly from real GitHub issues and pull requests across 12 widely used Python projects. 
Its tasks demand realistic, multi-file, multi-function modifications, offering a rigorous testbed for evaluating automated program repair (APR) systems.

Despite these advances, state-of-the-art models still fall short. 
Even sophisticated LLMs like Claude-2 have solved less than 2\% of SWE-bench tasks \cite{jimenez2023swebench}. 
More recent frameworks such as MAGIS \cite{tao2024magis} introduced multi-agent collaboration and achieved up to 13.94\% resolution rates, an order-of-magnitude improvement over GPT-4 baselines. 
Nonetheless, challenges remain: performance declines sharply as issue complexity grows, precise localization of edits remains unreliable, and generalization across repositories is inconsistent. 
These limitations underscore the necessity for new retrieval and localization strategies that can scale to repository-level complexity.

To address this gap, we propose \toolnospace, a hybrid retrieval approach designed to balance recall with conciseness in APR workflows. 
Our method integrates two complementary strategies: (1) codebase retrieval, where issue descriptions are matched against repository contents to identify potentially relevant files, and (2) history-based retrieval, where past issues and their associated edits guide candidate file selection. 
By merging and reranking candidates from both strategies, \toolnospace\ produces a concise yet comprehensive set of target files. 
Experiments on SWE-bench show that \toolnospace\ achieves higher recall without significantly increasing the number of retrieved files, thereby enabling more effective downstream APR.

In summary, this work contributes a novel retrieval framework tailored for large-scale repositories, motivated by the shortcomings of existing approaches and validated on the SWE-bench benchmark. 
By focusing on retrieval precision and efficiency, our framework moves one step closer to making repository-level automated program repair more practical and scalable.

\begin{table}[ht]
\centering
\begin{tabular}{lcc}
\hline
\textbf{Method} & \textbf{Model} & \textbf{\% Resolved} \\
\hline
TRAE             & Claude 4 Sonnet & 75.20 \\
Refact.ai Agent  & Claude 4 Sonnet & 74.40 \\
Moatless Tools   & Claude 4 Sonnet & 70.80 \\
OpenHands        & Claude 4 Sonnet & 70.40 \\
SWE-agent        & Claude 4 Sonnet & 66.60 \\
\hline
SWE-Fixer        & Qwen-2.5-7b     & 32.80 \\
\hline
\end{tabular}
\caption{Performance comparison on SWE-bench Verified leaderboard.}
\label{tab:swebench-leaderboard}
\end{table}

\section{Empirical Study}
The empirical study was conducted on the SWE-Bench benchmark, which consists of 268 real-world software engineering issues from popular Python repositories. 
The evaluation focused on analyzing instances where TRAE \cite{zhang2024trae}, an Automated Program Repair (APR) agent, failed to resolve issues, providing insights into the challenges and limitations of current approaches.

\subsection{Key Findings}
\subsubsection{Failure Analysis Across Repositories}

The study identified \textit{django/django} as the most challenging repository, with 23 instances where all agents failed. 
This was followed by \textit{sympy/sympy} (9 instances) and \textit{astroid/astroid} (8 instances). 
The concentration of failures in specific repositories suggests that certain codebases present systematic challenges for automated repair approaches, possibly due to their complexity, size, or architectural patterns.

\subsubsection{Difficulty Level Distribution}

The analysis revealed a strong correlation between task difficulty and agent failure rates (Figure ~\ref{fig:emp_1}a):
\begin{itemize}
    \item \textbf{>4 hours tasks}: 66.7\% failure rate across all agents, indicating that complex, time-intensive issues remain largely unsolvable
    \item \textbf{1-4 hours tasks}: 42.9\% failure rate, showing moderate difficulty
    \item \textbf{15 min - 1 hour tasks}: 12.3\% failure rate
    \item \textbf{<15 min fix tasks}: Only 3.1\% failure rate
\end{itemize}

This distribution demonstrates that current APR systems struggle disproportionately with tasks requiring deeper understanding and more extensive code changes.

\subsubsection{Temporal Patterns}

The year distribution (Figure ~\ref{fig:emp_1}a) shows that failure instances peaked during 2020-2021 (approximately 13 instances each year), with notable occurrences throughout 2019-2023. 
The 2018 baseline of 3 instances was relatively low. 
This temporal pattern may reflect the evolution of issue complexity in mature codebases or changes in how issues are reported and categorized.

\subsubsection{Error Type Patterns}

The error frequency analysis (Figure ~\ref{fig:emp_1}b) across repositories revealed:

\begin{itemize}
    \item \textbf{TypeError} dominated in \textit{sphinx-doc} (nearly 100 occurrences), suggesting type-related issues are particularly prevalent in documentation-focused codebases
    \item \textbf{AssertionError} was common across multiple repositories (\textit{django}, \textit{sympy}, \textit{pydata}, \textit{sphinx-doc}), indicating test failures and expectation mismatches
    \item \textbf{AttributeError} appeared frequently in \textit{django} and \textit{sphinx-doc}
    \item \textbf{KeyError} was notably present in \textit{django}
    \item \textbf{ValueError} showed modest frequency across \textit{astropy} and \textit{django}
\end{itemize}

The error distribution suggests that different repositories exhibit distinct failure patterns, likely reflecting their domain-specific characteristics and coding patterns.

\subsubsection{Failure Count Distribution}

\begin{figure*}[ht]
    \centering
    \includegraphics[width=0.9\textwidth]{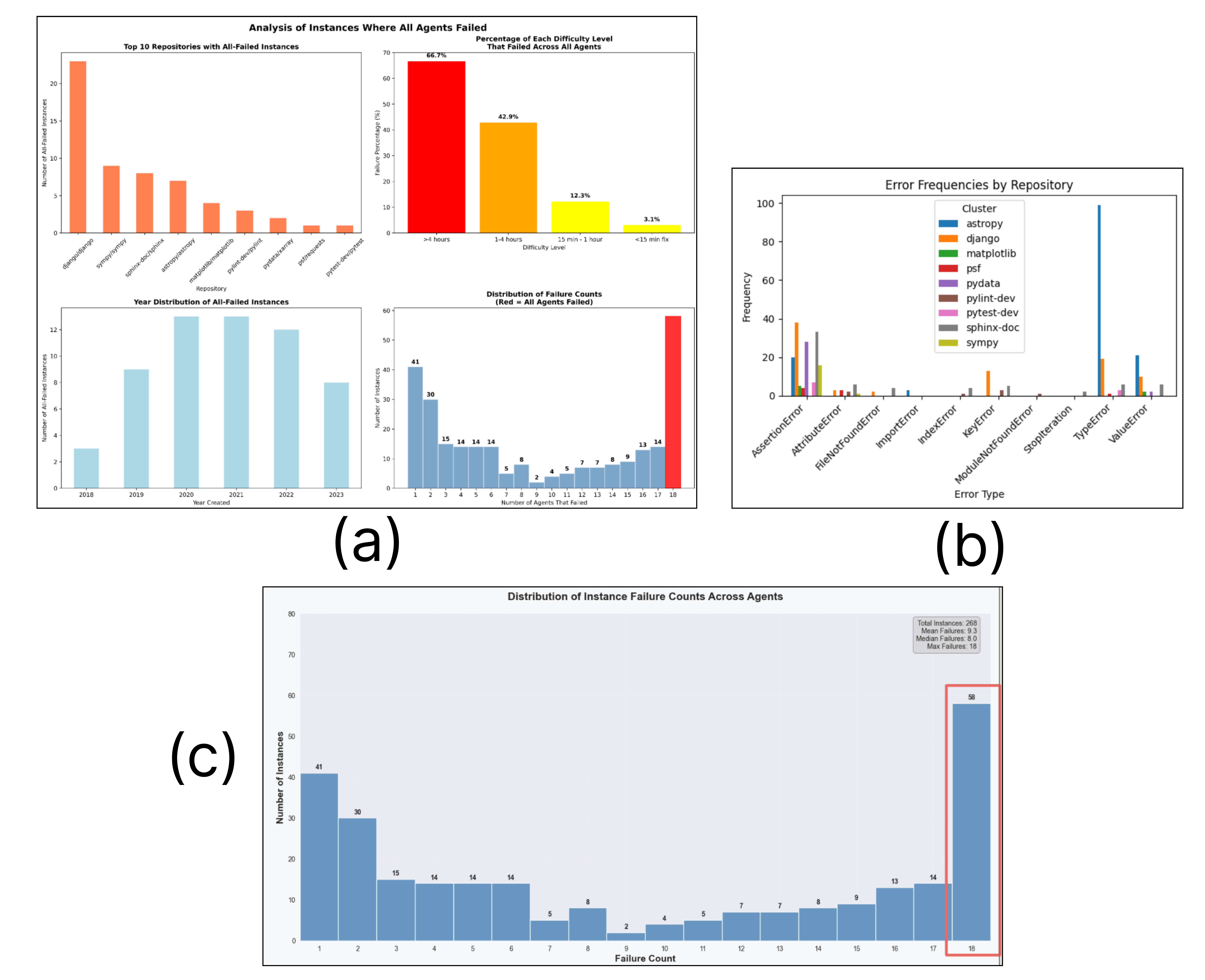}
    \caption{(a) Overview of failure patterns across repositories, difficulty levels, time periods, and failure counts. Top-left shows the top 10 repositories with the most all-failed instances. Top-right displays the percentage of each difficulty level that failed across all agents. Bottom-left presents the temporal distribution of all-failed instances from 2018 to 2023. Bottom-right shows the distribution of failure counts, with the red bar highlighting 58 instances where all 18 agents failed. (b) Error frequencies by repository across different error types. The chart reveals repository-specific error patterns, with TypeError dominating in sphinx-doc, AssertionError prevalent across multiple repositories, and varying frequencies of AttributeError, KeyError, and ValueError across different codebases. (c) Distribution of instance failure counts across agents. The histogram shows the number of instances for each failure count (1-18 agents). The red-highlighted bar at failure count 18 represents 58 instances where all agents failed, indicating particularly challenging cases in the benchmark.}
    \label{fig:emp_1}
\end{figure*}

A critical finding shown in the distribution analysis (Figure ~\ref{fig:emp_1}c) reveals:

\begin{itemize}
    \item 41 instances had only 1 agent failure
    \item 30 instances had 2 agent failures
    \item A long tail of instances with varying failure counts
    \item \textbf{58 instances had 18 agent failures} (highlighted in red), representing cases where all evaluated agents completely failed
\end{itemize}

The mean failure count was 9.3, with a median of 8.0 and maximum of 18, indicating substantial variability in task difficulty across the benchmark.

\subsection{Implications}

These findings underscore several challenges for the proposed hybrid retrieval approach:

\begin{enumerate}
    \item \textbf{Repository-specific adaptation}: The concentration of failures in specific repositories suggests that retrieval strategies may need to be tailored to different codebase characteristics
    
    \item \textbf{Scalability to complex tasks}: The high failure rate for >4 hour tasks indicates that improved file retrieval alone may be insufficient; deeper semantic understanding and multi-file reasoning capabilities are needed
    
    \item \textbf{Error-type awareness}: The diverse error patterns suggest that incorporating error-type information into the retrieval strategy could improve the identification of relevant files
    
    \item \textbf{Historical learning opportunity}: The 58 universal failure cases represent particularly valuable learning opportunities for improving both retrieval and repair strategies, as they likely contain common patterns that current approaches systematically miss
\end{enumerate}

Taken together, these findings point to file localization as a persistent bottleneck in automated program repair. 
Failures often arise not only from complex or long-duration tasks but also from the inability of existing systems to consistently identify the correct files for modification. 
Improving the retrieval stage becomes critical: retrieving too many files overwhelms the model with noise, while missing the true file leads to guaranteed failure. 
Motivated by this, our methodology focuses on enhancing file retrieval strategies, combining both repository-level and history-based signals to better balance recall and conciseness.

\section{Methodology}
\subsection{Traditional File Retrieval Approaches}
When an issue is reported in a large software repository, the first step in automated resolution is to identify the relevant files where code modifications are required. 
A widely adopted approach is to treat the issue description as a natural language query and retrieve candidate files from the entire codebase. 
Classical information retrieval techniques such as BM25 ~\cite{robertson2009probabilistic} are employed to rank files by their textual similarity to the issue description. 
The system then selects the top-$k$ files as the context for subsequent code generation or patch suggestion.

This retrieval paradigm is also reflected in the SWE-bench benchmark \cite{jimenez2023swebench}, where the baseline relies on BM25 to map issue descriptions to potentially relevant files. 
The retrieved files are then fed into large language models (LLMs) as context, with the expectation that the true locus of change is among them. 
While this method scales to large repositories with thousands of files, it introduces significant noise: models are frequently distracted by irrelevant context and fail to localize the exact site of modification.

\begin{figure}[]
\centering
\includegraphics[width=\columnwidth]{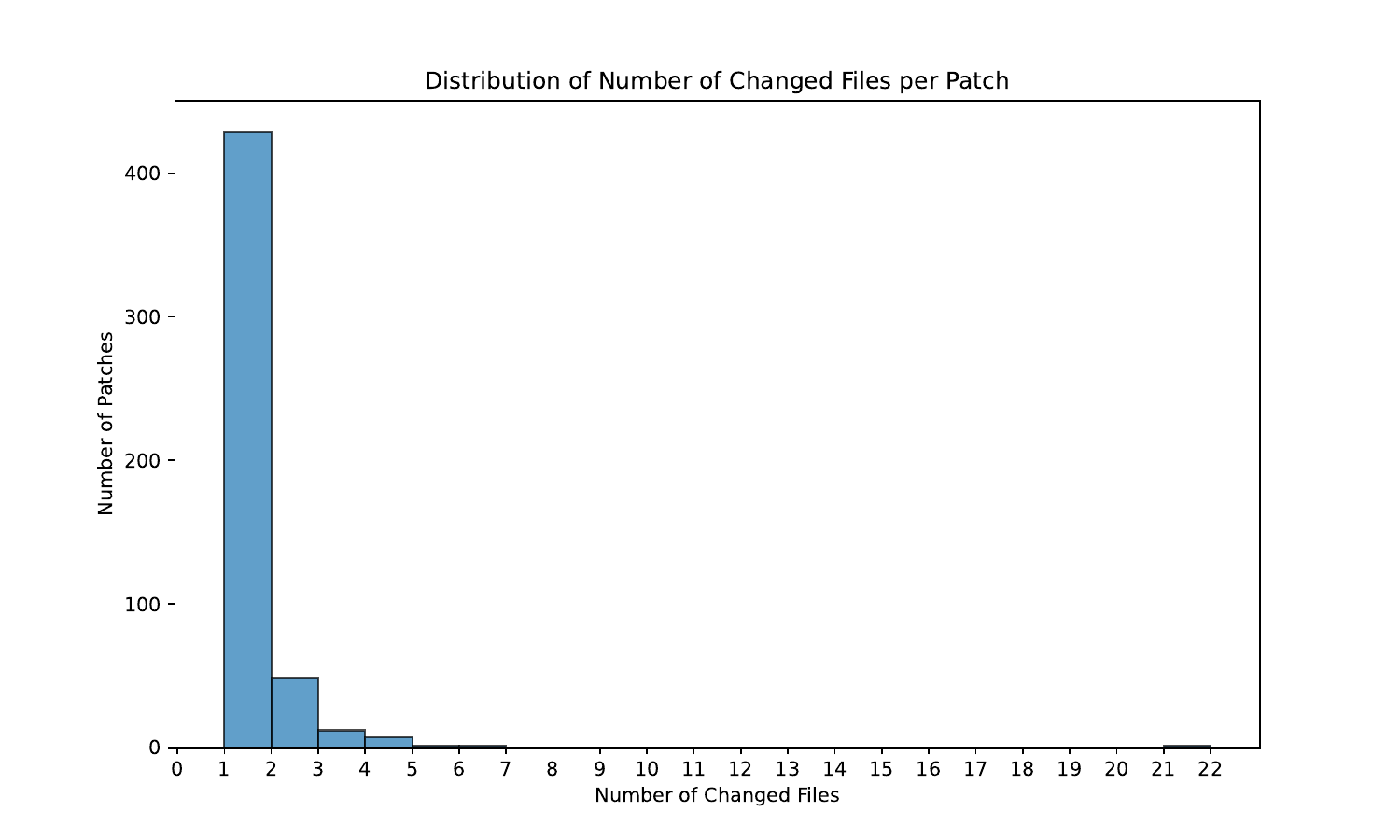}
\caption{Frequency distribution plot showing how many files were modified in the patches included in the SWE-bench-verified dataset. 
Our empirical analysis reveals that in more than 80\% of cases, only a single file actually requires modification. 
And in all of the patches, less than 10 files were modified.}
\label{fig_fig0}
\end{figure}

To better understand the distribution of code edits, we analyzed the SWE-bench-verified dataset, which contains 500 human-validated issue--patch pairs. 
Our analysis revealed that in more than 400 cases (over 80\% of the dataset), the resolution required changes in only a single file, as illustrated in Figure \ref{fig_fig0}. 

This finding underscores a key limitation of traditional retrieval approaches: although they retrieve a set of candidate files, in the vast majority of cases, only one file actually requires modification. 
Therefore, the retrieval stage must not only recall the correct file but also avoid overwhelming the downstream LLM with extraneous context. 
This motivates our investigation into hybrid retrieval strategies.

\subsection{Proposed Hybrid Retrieval Strategy}
While traditional retrieval methods rely solely on matching the issue text to the codebase, we propose a hybrid retrieval framework that integrates information from both verified and unverified instances of the SWE-bench dataset. 
The goal is to more accurately localize the files that require modification while minimizing the inclusion of irrelevant context.

\subsubsection{Leveraging Verified and Unverified Datasets}
We consider the SWE-bench-verified subset (500 human-validated issue--patch pairs) as the gold standard for evaluation, while treating the remainder of SWE-bench as an unverified pool of auxiliary data. 
When a new issue is presented, we first retrieve semantically similar issues from the unverified pool using a sentence transformer \cite{reimers2019sentencebert} model. 
From this step, we select the top-10 most relevant issues and extract their corresponding patch files. 
We consider 10 as a safe number, because Figure~\ref{fig_fig0} provides us with the evidence that all the patches in the verified dataset required less than 10 files to modify.
The files modified in these patches are then aggregated and scored according to their frequency and patch relevance, producing a candidate set of files likely to be relevant.

\subsubsection{Parallel BM25 Retrieval from the Codebase}
In parallel, we apply the traditional BM25-based retrieval directly on the target repository. 
This process computes the similarity between the input issue text and all files in the repository, selecting the top-$k$ candidate files. 
Thus, we obtain two ranked lists of candidate files: one derived from issue--patch similarity in the unverified dataset, and another derived from direct issue--file similarity within the target codebase.

\subsubsection{Re-ranking with Hybrid Scoring}
After obtaining candidate files from both retrieval streams, we re-rank the union of their results to select the most relevant top-$k$ files. 
Each stream produces a ranked list of \((\texttt{docid}, \texttt{score})\) tuples with method-specific scoring. 
To make scores comparable, we apply min--max normalization \emph{per method, per instance}:
\[
s'_{m}(f) = \frac{s_{m}(f) - \min(s_{m})}{\max(s_{m}) - \min(s_{m}) + \epsilon},
\]
where $m \in \{\text{SentenceTransformer}, \text{BM25}\}$, $s_{m}(f)$ is the raw score for file $f$, and $\epsilon$ is a small constant for stability.
We then compute a hybrid score
\[
H(f) = \alpha \cdot s'_{\text{ST}}(f) + (1-\alpha) \cdot s'_{\text{BM25}}(f),
\]
where $s'_{\text{ST}}(f)$ comes from a Sentence Transformer retriever (we use \texttt{all-mpnet-base-v2 \cite{song2020mpnet, reimers2021allmpnet}} indexed with FAISS over the unverified pool) and $s'_{\text{BM25}}(f)$ from sparse retrieval over the target codebase. 
Files missing from one stream receive a normalized score of $0$ for that stream. 
We sort candidates by $H(f)$ and keep the top-$k$.

To study sensitivity, we sweep $\alpha \in [0,1]$ and $k \in \{1,\dots,10\}$, computing recall@$k$ on \textit{SWE-bench-verified}. 
This exposes the trade-off between dense semantic similarity and sparse lexical matching; intermediate $\alpha$ values typically yield the most robust recall. 
The re-ranked top-$k$ files are then provided as context to the downstream LLM for patch generation.

\begin{figure}[]
    \centering
    \includegraphics[width=\linewidth]{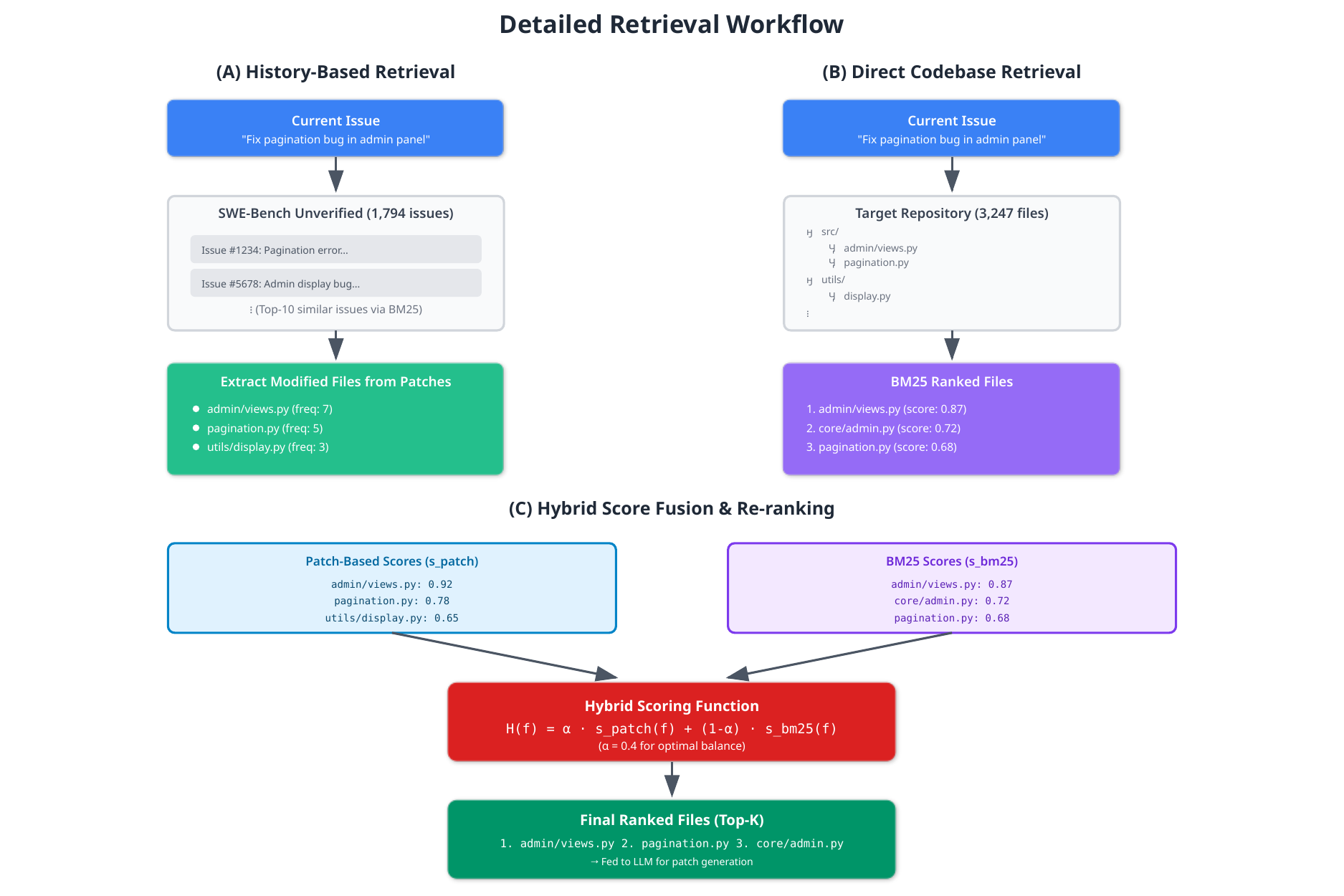}
    \caption{Overview of our proposed hybrid retrieval workflow. 
    \textbf{(A) History-Based Retrieval:} The current issue is compared with the SWE-bench-unverified dataset using sentence transformer to retrieve the top-10 most similar issues. 
    From their associated patches, the modified files are extracted, and scored. 
    \textbf{(B) Direct Codebase Retrieval:} In parallel, the current issue is matched against the entire target repository using BM25, producing a ranked list of candidate files by textual relevance. 
    \textbf{(C) Hybrid Score Fusion \& Re-ranking:} Scores from both retrieval streams are min--max normalized and combined via the hybrid scoring function $H(f)$. 
    The fused list is re-ranked to produce the top-$k$ files, which are supplied as input context to the LLM for patch generation.}
    
    \label{fig:workflow}
\end{figure}

The full workflow of our pipeline has been illustrated in Figure~\ref{fig:workflow}.
 
\section{Experiments and Analysis}
\subsection{Experimental Setup}
To evaluate the effectiveness of our retrieval strategies, we conduct experiments on the SWE-bench dataset \cite{jimenez2023swebench}. 
We partition the dataset into two subsets: \textit{SWE-bench-verified}, which consists of 500 human-validated issue--patch pairs, 
and \textit{SWE-bench-unverified}, which consists of the remaining 1,794 instances. 
Our retrieval experiments are primarily benchmarked on the verified subset, as it serves as the gold standard for correctness.

We implement three retrieval baselines: 
\begin{enumerate}
    \item \textbf{BM25}: a probabilistic sparse retrieval method widely used in SWE-bench baselines.
    \item \textbf{TF--IDF \cite{salton1988termweighting}}: a classical term-weighting scheme for ranking files by cosine similarity with the issue description.
    \item \textbf{Sentence Transformer}: a dense retrieval model (we use the \texttt{all-mpnet-base-v2} variant) 
    that encodes both issue text and file contents into embeddings, and ranks files by semantic similarity.

\end{enumerate}

Each method takes as input the issue description and retrieves the top-$k$ files from the codebase. 
We evaluate retrieval quality using \textit{recall}---the percentage of gold (ground-truth) edited files contained within the retrieved set. 
This metric directly captures how effectively a retrieval strategy localizes the true locus of change.

\subsection{Baseline Retrieval Performance}
\begin{figure}[]
    \centering
    \includegraphics[width=\linewidth]{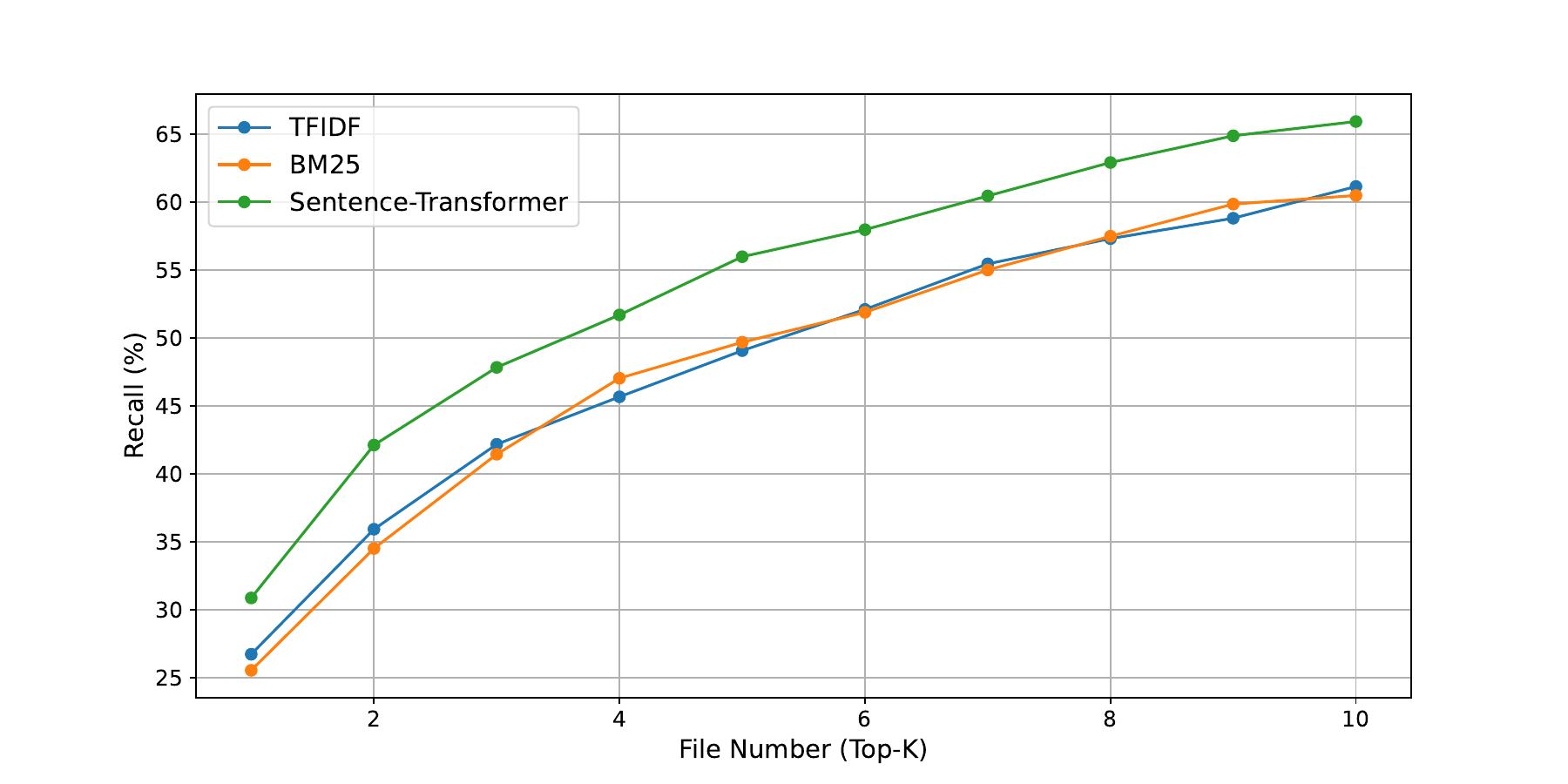}
    \caption{Comparison of recall scores for BM25, TF--IDF, and Sentence Transformer retrievers on SWE-bench-verified. 
    The $x$-axis denotes the number of retrieved files ($k$), while the $y$-axis shows recall. 
    The sentence transformer demonstrates the highest recall across all retrieval depths, highlighting its advantage in bridging semantic gaps between issue descriptions and code.}
    \label{fig:baseline-recall}
\end{figure}

Figure~\ref{fig:baseline-recall} presents recall performance across different values of $k$. 
We observe that the sentence transformer consistently outperforms BM25 and TF--IDF across all settings, demonstrating the benefit of semantic embeddings in capturing relationships between issue descriptions and source code. 
BM25 performs competitively at low $k$, but its recall saturates more quickly. 
TF--IDF trails both methods, indicating limitations in handling natural language and code vocabulary mismatch.

\subsection{Hybrid Retrieval with Score Fusion}
Building on these baselines, we evaluate our proposed hybrid retrieval approach, which integrates \textit{issue--patch similarity} from the unverified subset with direct BM25 retrieval on the codebase. 
We combine the two sources of evidence by normalizing scores and computing a hybrid relevance score:
\[
H(f) = \alpha \cdot s'_{\text{ST}}(f) + (1-\alpha) \cdot s'_{\text{BM25}}(f),
\]
where $s'_{\text{ST}}(f)$ comes from a Sentence Transformer retriever (we use \texttt{all-mpnet-base-v2} indexed with FAISS over the unverified pool) and $s'_{\text{BM25}}(f)$ from sparse retrieval over the target codebase. 

\begin{figure}[]
    \centering
    \includegraphics[width=\linewidth]{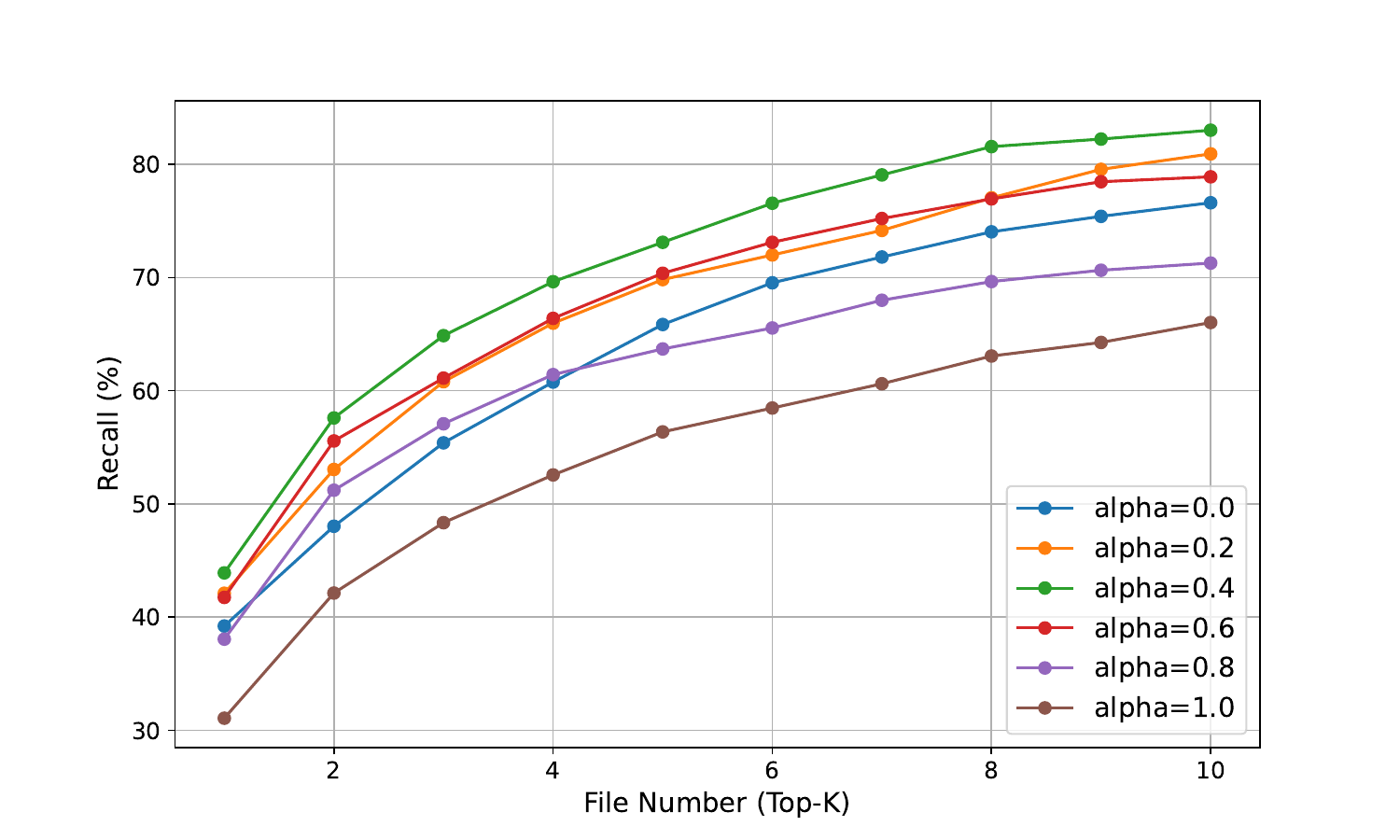}
    \caption{Recall scores of the proposed hybrid retrieval strategy on SWE-bench-verified for different values of $\alpha$. 
    When $\alpha=0$, the system reduces to pure BM25 retrieval; when $\alpha=1$, it relies solely on issue--patch similarity. 
    Hybrid configurations with $\alpha$ between $0.4$ and $0.6$ achieve the highest recall, demonstrating the benefit of balancing patch-derived evidence with direct textual similarity.}
    \label{fig:hybrid-recall}
\end{figure}

Figure~\ref{fig:hybrid-recall} shows recall results for varying $\alpha$. 
We find that hybrid retrieval improves robustness across different $k$ values, with intermediate $\alpha$ values (e.g., $\alpha=0.4$ to $0.6$) yielding the best overall performance. 
This confirms that combining historical patch-level information with direct code retrieval allows for more precise localization.

\section{Related Work}
\begin{figure}[]
    \centering
    \includegraphics[width=\linewidth]{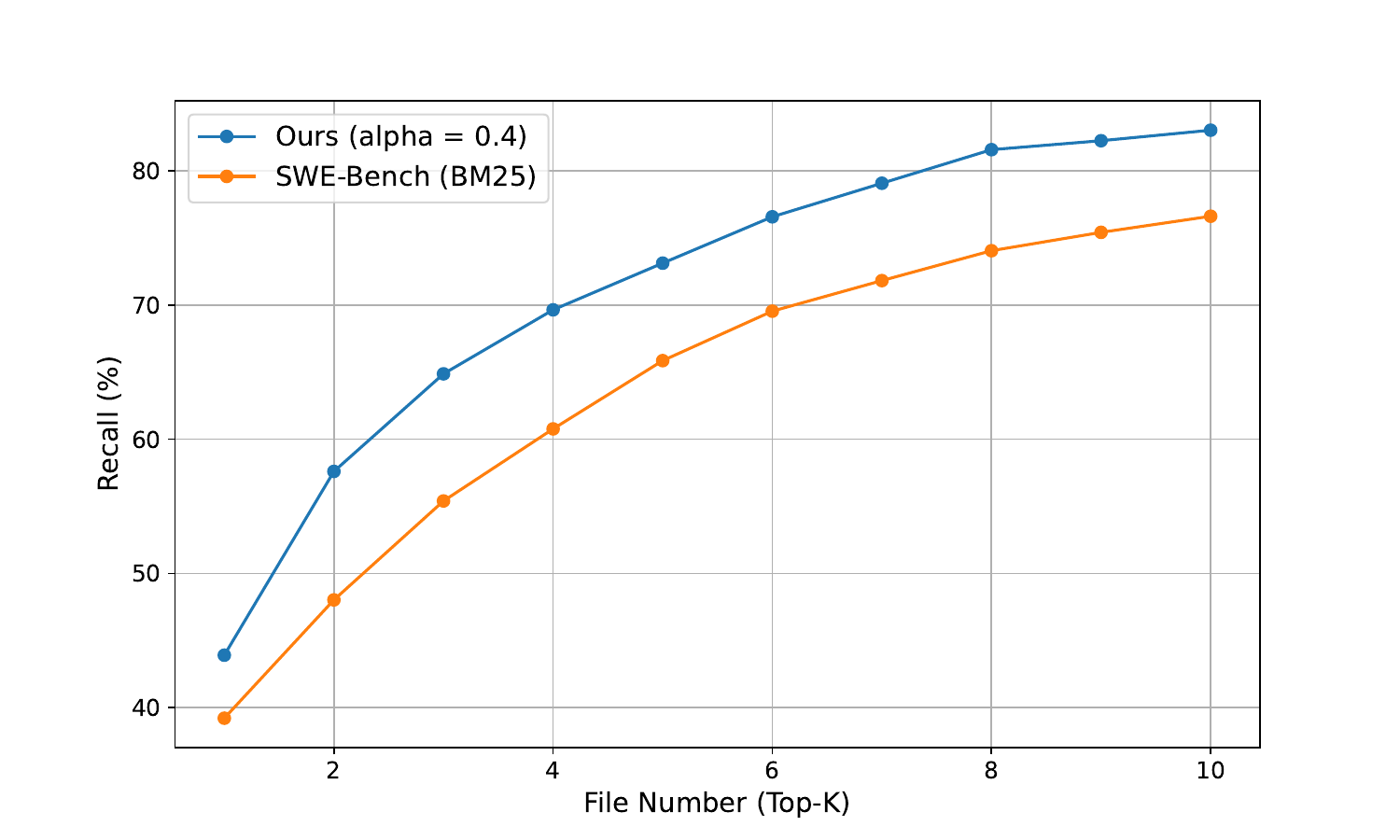}
    \caption{Recall@K comparison between our retrieval methods and the BM25 baseline used in MAGIS. 
    The $x$-axis denotes the number of retrieved files ($k$), while the $y$-axis shows recall percentage. 
    Our sentence transformer retriever, combined with hybrid scoring ($\alpha = 0.4$), achieves consistently higher recall, highlighting the effectiveness of semantic and patch-augmented retrieval compared to sparse BM25 retrieval alone.}
    \label{fig:magis-vs-ours}
\end{figure}

MAGIS~\cite{tao2024magis} introduces a multi-agent framework for GitHub issue resolution, where specialized agents collaborate to identify relevant files, generate candidate patches, and validate solutions. 
A key contribution of MAGIS lies in its emphasis on repository-scale context handling and multi-step coordination, addressing some of the limitations of single-agent retrieval and patching systems. 
Despite achieving substantial improvements over baseline methods (e.g., 13.94\% resolution rate compared to 1.96\% for Claude 2 on SWE-bench), MAGIS still struggles with precise file localization, particularly in repositories with large and complex dependency structures. 
Their reported recall curves (Figure~\ref{fig:magis-vs-ours}) highlight that while BM25 provides a reasonable sparse retrieval baseline, performance saturates quickly and leaves significant room for enhancement.

In contrast, our work explores hybrid retrieval strategies that fuse semantic retrieval models with patch-derived signals from the unverified portion of SWE-bench. 
As shown in Figure~\ref{fig:magis-vs-ours}, 
our sentence transformer model achieves consistently higher recall than BM25 across all retrieval depths. 
Moreover, by incorporating hybrid score fusion (with $\alpha = 0.4$), we surpass the BM25 baseline used in MAGIS, demonstrating that lightweight hybrid retrieval can yield complementary benefits to multi-agent systems. 
This comparison underscores that while MAGIS focuses on agent collaboration, our method advances the retrieval stage itself, which is a critical bottleneck for effective LLM-based issue resolution.

\section{Conclusion}
In this work, we addressed the challenge of file retrieval for automated program repair in large software repositories. 
Through an empirical study on the SWE-bench dataset, we confirmed that traditional retrieval approaches such as BM25, while widely used, introduce significant noise by retrieving multiple irrelevant files despite the fact that over 80\% of SWE-bench-verified tasks require edits in only a single file. 
To mitigate this limitation, we proposed \textit{\tool}, a hybrid retrieval strategy that integrates direct codebase retrieval with history-based retrieval from past issue--patch pairs. 
By normalizing and fusing scores from both sources, our method balances precision and recall, ensuring that downstream LLMs are provided with more concise yet comprehensive contexts.

Our experiments demonstrated that dense retrieval with sentence transformers outperforms sparse methods like BM25 and TF--IDF, and that the hybrid score fusion further improves recall across different retrieval depths. 
Compared with existing multi-agent frameworks such as MAGIS, our approach advances the retrieval stage itself, achieving consistently higher recall with a lightweight, modular design that can complement more complex agent-based systems.

Looking forward, we envision extending our framework in several directions. 
First, adaptive weighting strategies for score fusion could dynamically tune $\alpha$ based on issue complexity. 
Second, integrating structural signals from abstract syntax trees (ASTs) or dependency graphs may provide more fine-grained localization. 
Finally, combining our retrieval improvements with generation- and validation-focused systems has the potential to substantially raise end-to-end resolution rates on SWE-bench and beyond. 
By strengthening the retrieval stage, our work takes a step towards making automated program repair more practical, scalable, and reliable in real-world software engineering contexts.

\bibliography{anthology,custom}
\bibliographystyle{acl_natbib}

\end{document}